\documentclass[11pt]{article}
\usepackage{geometry}
\usepackage{ifpdf}
\ifpdf
  \usepackage[pdftex]{graphicx}
  \usepackage{epstopdf}
  \usepackage[pdftex,bookmarks=false,colorlinks=true,pdfstartview=FitBV,linkcolor=blue,citecolor=blue,urlcolor=blue]{hyperref}
\else
  \usepackage{graphicx}
  \usepackage{epsfig,psfrag}
  \psdraft\psfull
\fi
\usepackage{graphicx, enumerate}
\usepackage{cite}
\usepackage{subfigure, epsfig, subeqnarray}
\usepackage[sub,ovp]{psfragx}
\usepackage{srcltx}
\usepackage{subeqnarray}
\usepackage{latexsym,amssymb,amsmath,amsfonts,amscd,mathrsfs}
\usepackage{pdfpages,pdfsync}
\usepackage{color}
\usepackage{comment}
\usepackage{float}
\floatstyle{plaintop}
\restylefloat{table}

\psdraft \psfull

\newcommand{\dd}{\mathrm{d}}
\newcommand{\R}{{\mathbb{R}}}

\newcommand{\F}{\mathscr{F}}
\newcommand{\prob}{\mathbb{P}}
\newcommand{\probb}{\mathbb{Q}}
\newcommand{\E}{\mathbb{E}}
\newcommand{\UU}{\mathscr{U}[0,T]}
\newcommand{\tr}{\mathrm{tr}}
\newcommand{\Dti}{\Delta t_i}
\newcommand{\DWi}{\Delta W_i}

\newcommand{\Zti}{Z_{t_i}}
\newcommand{\Yti}{Y_{t_i}}

\newcommand{\Xt}{\tilde{X}}
\newcommand{\Yt}{\tilde{Y}}
\newcommand{\Zt}{\tilde{Z}}
\newcommand{\Xstx}{X_s^{t,x}}
\newcommand{\Ystx}{Y_s^{t,x}}
\newcommand{\Zstx}{Z_s^{t,x}}

\newtheorem{Theorem}{Theorem}

\newtheorem{assumption}{Assumption}

\title{\LARGE \bf
	Learning Optimal Control via Forward and Backward Stochastic Differential Equations
}

\date{}
\author{
	Ioannis Exarchos  \textnormal{and} Evangelos A. Theodorou 
	\\
	Department of Aerospace Engineering\\
	Georgia Institute of Technology\\
	Atlanta, GA 30332 \\
	\texttt{exarchos@gatech.edu},  \texttt{evangelos.theodorou@ae.gatech.edu}
}

\begin{document}

	\maketitle
	\thispagestyle{empty}
	\pagestyle{empty}

	\begin{abstract}
		In this paper we present a novel sampling-based numerical scheme designed to solve a certain class of stochastic optimal control problems, utilizing forward and backward stochastic differential equations (FBSDEs). By means of a nonlinear version of the Feynman-Kac lemma, we obtain a probabilistic representation of the solution to the nonlinear Hamilton-Jacobi-Bellman equation, expressed in the form of a decoupled system of FBSDEs. This system of FBSDEs can then be simulated by employing linear regression techniques. To enhance the efficiency of the proposed scheme when treating more complex nonlinear systems, we then derive an iterative modification based on Girsanov's theorem on the change of measure, which features importance sampling. The modified scheme is capable of learning the optimal control without requiring an initial guess. We present simulations that validate the algorithm and demonstrate its efficiency in treating nonlinear dynamics.
	\end{abstract}

	\section{Introduction}
	The problem of obtaining an optimal control in a stochastic setting is typically associated with the solution of a generally nonlinear second-order partial differential equation (PDE), known as the Hamilton-Jacobi-Bellman (HJB) equation. Different methods can be distinguished based on whether they seek a solution over the entire domain, or locally around a nominal system trajectory. In the first case, several attempts have been made to address the inherent difficulty in solving such nonlinear PDEs, as well as the curse of dimensionality \cite{Mitchell2003,Aguilar2014,Horowitz2014,Horowitz2014a,Gorodetsky2015}. The latter case, on the other hand, includes methods such as Stochastic Differential Dynamic Programming \cite{Todorov2005,Theodorou2010}, which is based on linearization of the dynamics and a quadratic approximation of the cost functions around nominal trajectories, as well as sampling-based methods, which are distinguished for their ability to accommodate scalable iterative schemes. Various sampling-based methods appear in the literature under the names Path Integral control (PI) \cite{Kappen2005,Theodorou2010a}, Kullback Leibler (KL) control, or Linearly Solvable Control \cite{Dvijotham2012,Todorov2009}.

	The fundamental characteristic of all aforementioned sampling-based methods is that they rely on the exponential transformation of the Value function \cite{Theodorou2015}, and the linear version of the Feynman-Kac lemma \cite{Karatzas1991}.  Under the exponential transformation, and by introducing certain restrictions between control authority and stochasticity, there exists a direct relationship  between the Hamilton-Jacobi-Bellman PDE and the backward Chapman-Kolmogorov PDE.  The latter, being a linear PDE, enables the use of the linear Feynman-Kac formula, which relates certain linear backward PDEs to forward stochastic differential equations (SDEs). Thus, the corresponding optimal control problem can be solved using forward sampling.  While forward sampling-based methods exhibit several advantages against traditional methods of stochastic control, such as the mild conditions on the differentiability of the cost and the stochastic dynamics, there are also some key disadvantages which pertain to the nature of the exponential transformation. In particular, the effect of the exponential transformation can be identified as the mapping of the value function $v(t,x)$, which has range $[0, \infty)$, to the desirability function $\psi(t,x)$, whose range is $ (0, 1] $.  This mapping leads to a drastic reduction in the ability to distinguish states with high cost (low desirability) from states with low cost (high desirability).  This issue has been partially addressed with renormalization of the trajectory cost \cite{Theodorou2010}. Finally, while the necessary constraint introduced between control authority and stochasticity  can lead to symmetry breaking phenomena and delayed decision \cite{Kappen2005a}, it is a rather restrictive assumption whenever applications to engineered systems are considered. 
	
	In this paper we present a learning control algorithm which capitalizes on the innate relationship between certain \textit{nonlinear} PDEs and Forward and Backward SDEs, demonstrated by a \textit{nonlinear} version of the Feynman-Kac lemma.  By means of this lemma,  we obtain a probabilistic representation of the solution to the  nonlinear HJB PDE, expressed in the form of a system of decoupled FBSDEs. This system of FBSDEs can then be simulatedby  employing linear regression techniques. We wish to highlight the difference between the proposed approach and already existing sampling-based formulations: our approach addresses directly the nonlinear PDE, while the latter make use of the exponential transformation, which under certain conditions yields a linear PDE problem, and then use forward sampling to address that linear problem. Thus, the herein proposed framework relaxes these restrictive conditions. To enhance the efficiency of the proposed scheme when treating more complex nonlinear systems, we then derive an iterative algorithm based on Girsanov's theorem on the change of measure, which features importance sampling. The proposed scheme is capable of learning the optimal control without requiring an initial guess. We present simulations that validate the algorithm and demonstrate its efficiency in treating nonlinear dynamics.
	
	\section{Problem Statement}\label{PF}
	Let $(\Omega,\F,\{\F_t\}_{t\geq 0},\prob)$ be a complete, filtered probability space on which is defined a $p$-dimensional standard Brownian motion $W_t$, such that $\{\F_t\}_{t\geq 0}$ is the natural filtration of $W_t$ augmented by all $\prob$-null sets. Consider the problem of minimizing the expected cost defined by the cost functional
	\begin{equation} \label{cost}
		J(\tau, x_\tau; u(\cdot))=\E \bigg[ g(x(T)) + \underset{\tau}{\overset{~~T}{\int}}q(t,x(t))+\frac{1}{2}u^{\top} (t)Ru(t) \dd t \bigg],
	\end{equation}
	associated with the stochastic controlled system which is represented by the It\^{o} stochastic differential equation (SDE)
	\begin{equation}\label{stdyn}
	\left \{
		\begin{aligned}
			&\dd x(t)=f(t,x(t))\dd t +G (t,x(t))u(t)\dd t + \Sigma(t,x(t))\dd W_t,\qquad t\in[\tau,T],\\& x(\tau)=x_\tau,
		\end{aligned}
			\right .
	\end{equation}
	where $T>\tau\geq 0$, $T$ is a fixed time of termination, $x \in \R^n$ is the state vector, $u \in \R^\nu$ is the control vector, $R$ is a $\nu \times\nu$ positive definite matrix, and $g:\R^n\to\R$, $q:[0,T]\times \R^n \to \R$, $f:[0,T]\times \R^n \to \R^n$, $G:[0,T]\times \R^n \to \R^{n\times \nu}$, and $\Sigma:[0,T]\times \R^n\to\mathbb{R}^{n \times p}$ are deterministic functions, that is, they do not depend explicitly on $\omega \in \Omega$.
	We assume that all standard technical conditions \cite{Yong1999} which pertain to the filtered probability space and the regularity of functions are met, in order to guarantee existence, uniqueness of solutions to (\ref{stdyn}), and a  well defined cost functional (\ref{cost}). These impose for example that the functions $g$, $q$, $f$, $G$ and $\Sigma$ are continuous w.r.t. time $t$ (in case there is explicit dependence), Lipschitz (uniformly in $t$) with respect to the state variables and satisfy standard growth conditions over the domain of interest. Furthermore, the square-integrable process $u:[0,T]\times\Omega \to U\subseteq \R^\nu$ is $\{\F_t\}_{t\geq 0}$-adapted (also called \textit{progressively measurable}), which essentially translates into the control input being non-anticipating, i.e.,  relying only on past and present information. We denote the set of all admissible $U$-valued functions as $\UU$. For any given initial condition $(\tau,x_\tau)$, we wish to minimize (\ref{cost}) under all admissible functions $u(\cdot) \in \UU$. We define the Value function $V$ as
	\begin{equation}\label{Val}
	\left \{
		\begin{aligned}
			V(\tau,x_\tau)&=\inf_{u(\cdot)\in \UU}J(\tau, x_\tau;u(\cdot)),~  (\tau,x_\tau) \in [0,T)\times \R^n \\
			 \quad V(T,x)&=g(x), \quad  x\in \R^n.
		\end{aligned}
		\right .
	\end{equation}
	By applying the stochastic version of Bellman's principle of optimality, it is shown \cite{Yong1999,Fleming2006} that if the Value function is in $C^{1,2}([0,T]\times\R^n)$, then it is a solution to the following terminal value problem of a second-order partial differential equation, known as the Hamilton-Jacobi-Bellman (HJB) equation, which, for the problem at hand, and suppressing function arguments for notational compactness, takes the form
	\begin{equation}\label{HJB}
	\left \{
	\begin{aligned}
	&v_t+\inf_{u\in U}\bigg\{\frac{1}{2} \tr (v_{xx}\Sigma\Sigma^{\top}) + v_x^{\top} (f+Gu)
	+q+\frac{1}{2}u^{\top}Ru\bigg\}=0, \quad(t,x)\in [0,T)\times \R^n, \\
	&v(T,x)=g(x), \quad x\in \R^n.
	\end{aligned}
	\right .
	\end{equation}
	where $v_x$ and $v_{xx}$ denote the gradient and the Hessian of $v$, respectively. The term inside the brackets is the Hamiltonian, and is denoted by $H$.  Note that this result can be extended to include cases where the Value function does not satisfy the smoothness condition. Then, if one also considers viscosity solutions of (\ref{HJB}), the Value function is proven to be a viscosity solution of (\ref{HJB}). Furthermore, the viscosity solution is equal to the classical solution, if a classical solution exists. For the chosen form of the cost integrand, and assuming that the optimal control lies in the interior of $U$, we may carry out the infimum operation by taking the gradient of the Hamiltonian with respect to $u$ and setting it equal to zero to obtain
	\begin{equation}
	\frac{\partial H}{\partial u}=0\quad \mathrm{or}\quad -Ru-G^{\top}(t,x)v_x(t,x)=0.
	\end{equation}
	Therefore, the optimal control is given by
	\begin{equation}\label{OC}
	u^*(t,x)=-R^{-1}G^{\top}(t,x)v_x(t,x), \qquad (t,x)\in [0,T]\times\R^n.
	\end{equation}
	Inserting the above expression back into the original HJB equation and suppressing again function arguments, we obtain the equivalent characterization
	\begin{equation}\label{HJB2}
	\left \{
	\begin{aligned}
	&v_t+\frac{1}{2}\tr (v_{xx}\Sigma\Sigma^{\top})+v_x^{\top}f+q-\frac{1}{2}v_x^{\top}GR^{-1}G^{\top}v_x =0, \quad (t,x)\in [0,T)\times \R^n,\\&v(T,x)=g(x),\qquad
	x\in \R^n.
	\end{aligned}
	\right .
	\end{equation}
	
	\section{A Feynman-Kac Type Representation Through FBSDEs}	\label{sec:Feynman_Kac}
	There is an innate relation between stochastic differential equations and second-order partial differential equations of parabolic or elliptic type. Specifically, solutions to a certain class of nonlinear PDEs  can be represented by solutions to forward-backward stochastic differential equations (FBSDEs), in the same spirit as demonstrated by the well-known Feynman-Kac formulas \cite{Karatzas1991} for linear PDEs. We begin by briefly reviewing FBSDEs.
	\subsection{The Forward and Backward Process}\label{FBSDEpresent}
	As a forward process we shall define the square-integrable, $\{\F_s\}_{s\geq 0}$-adapted process $X(\cdot)$\footnote{While $X$ is a function of $s$ and $\omega$, we shall use $X_s$ for notational brevity.}, which, for any given initial condition $(t,x) \in [0,T]\times\R^n$, satisfies the It\^{o} FSDE
	\begin{equation}\label{fsde}
	\left \{
	\begin{aligned}
	\dd X_s&=b(s,X_s)\dd s + \Sigma(s,X_s)\dd W_s, \qquad s\in[t,T],\\
	X_t&=x.
	\end{aligned}
	\right.
	\end{equation}
	The forward process (\ref{fsde}) is also called the \textit{state process} in the literature.  We shall denote the solution to the forward SDE (\ref{fsde}) as $\Xstx$, wherein $(t,x)$ are the initial condition parameters.
	
	In contrast to the forward process, the associated backward process is the square-integrable, $\{\F_s\}_{s\geq 0}$-adapted pair $(Y(\cdot),Z(\cdot))$ defined via a BSDE satisfying a terminal condition
	\begin{equation}\label{bsde}
	\left \{
	\begin{aligned}
	\dd Y_s&=-h(s,X_s,Y_s,Z_s)\dd s +Z_s^{\top}\dd W_s \qquad s\in[t,T],\\
	Y_T&=g(X_T).
	\end{aligned}
	\right.
	\end{equation}
	The function $h(\cdot)$ is called \textit{generator} or \textit{driver}. The solution is implicitly defined by the initial condition parameters $(t,x)$ of the FSDE since it obeys the terminal condition $g(X^{t,x}_T)$. We will similarly use the notation $\Ystx$ and $\Zstx$ to denote the solution for a particular initial condition parameter $(t,x)$ of the associated FSDE.
	
	While FSDEs have a fairly straightforward definition, in the sense that both the SDE and the filtration evolve forward in time, this is not the case for BSDEs. Indeed, since solutions to BSDEs need to satisfy a terminal condition, integration needs to be performed backwards in time in some sense, yet the filtration still evolves forward in time. It turns out \cite{Ma1999} that a terminal value problem involving BSDEs admits an adapted (i.e., non-anticipating) solution if we back-propagate the \textit{conditional expectation} of the process, that is, if we set $Y_s\triangleq\E[Y_s|\F_s]$.
	
	Notice that the FSDE does not depend on $Y_s$ or $Z_s$. Thus, the resulting system of FBSDEs is said to be \textit{decoupled}. If, in addition, the functions $b$, $\Sigma$, $h$ and $g$ are deterministic, in the sense that they do not depend explicitly on $\omega\in\Omega$, then the adapted solution $(Y,Z)$ exhibits the \textit{Markovian} property; namely, it can be written as deterministic functions of solely time and the state process \cite{ElKaroui1997}: 
	\begin{Theorem}  \textit{(The Markovian Property) --
			There exist deterministic functions $v(t,x)$ and $d(t,x)$\footnote{By abuse of notation, here $(t,x)$ are symbolic arguments of the functions $v$ and $d$, and not the initial condition parameters as in $(Y^{t,x}, Z^{t,x})$. Throughout this work, it should be clear from the context whether $(t,x)$ are to be understood as initial condition parameters or symbolic arguments.} such that the solution $(Y^{t,x}, Z^{t,x})$ of the BSDE (\ref{bsde}) is
			\begin{equation}\label{MarkovP}
			\Ystx=v(s,\Xstx),\quad \Zstx=\Sigma^{\top}(s,\Xstx)d(s,\Xstx),
			\end{equation}
			for all $s\in[t,T]$.}
	\end{Theorem}
	\subsection{The Nonlinear Feynman-Kac Lemma}
	We now proceed to state the nonlinear Feynman-Kac type formula, which links the solution of a class of PDEs to that of FBSDEs. Indeed, the following theorem can be proven \cite{Yong1999,ElKaroui1997,Ma1999} by an application of It\^{o}'s formula:
	\begin{Theorem} \textit{ (Nonlinear Feynman-Kac) --
	Consider the Cauchy problem
	\begin{equation}\label{Cauchy}
	\left \{
	\begin{aligned}
	&v_t+\frac{1}{2}\tr (v_{xx}\Sigma\Sigma^{\top})+ v_x^{\top}b(t,x) +h(t,x,v,\Sigma^{\top}v_x) =0,\quad (t,x)\in [0,T)\times \R^n,\\
	&v(T,x)=g(x), \quad x\in \R^n,
	\end{aligned}
	\right .
	\end{equation}
	wherein the functions $\Sigma$, $b$, $h$ and $g$ satisfy mild regularity conditions. Then (\ref{Cauchy}) admits a unique viscosity solution $v:[0,T]\times\R^n\to \R$, which has the following probabilistic representation:
	\begin{equation}\label{FKrep}
	v(t,x)=Y_t^{t,x}, \qquad \forall (t,x)\in [0,T]\times \R^n,
	\end{equation}
	where $(X(\cdot),Y(\cdot),Z(\cdot))$ 
	is the unique 
    adapted solution of the FBSDE system (\ref{fsde})-(\ref{bsde}). Furthermore, 
	\begin{equation}\label{FKbsde}
	(\Ystx,\Zstx)=\bigg(v(s,\Xstx),~\Sigma^{\top}(s,\Xstx)v_x(s,\Xstx)\bigg),
	\end{equation}
	for all $s\in[t,T]$, and if (\ref{Cauchy}) admits a classical solution, then (\ref{FKrep}) provides that classical solution.}
	\end{Theorem}

	A comparison of equations (\ref{HJB2}) and (\ref{Cauchy}) indicates that the nonlinear Feynman-Kac representation can be applied to the HJB equation given by (\ref{HJB2}) under a certain decomposability condition, stated in the following assumption:
	\begin{assumption}\label{GSigma}
		\textit{There exists a matrix-valued function $\Gamma:[0,T]\times\R^n\to\R^{p \times \nu}$ such that $G(t,x)=\Sigma(t,x)\Gamma(t,x)$ for all $(t,x)\in[0, T]\times \R^n$, satisfying the same mild regularity conditions.}
	\end{assumption}
	This assumption implies that the range of $G$ must be a subset of the range of $\Sigma$, and thus excludes the case of  a channel containing control input but no noise, although the converse is allowed.
	Under this assumption, and suppressing arguments for brevity,
	the HJB equation given by (\ref{HJB2}) becomes
	\begin{equation}\label{HJB3}
		\left \{
		\begin{aligned}
			&v_t+\frac{1}{2}\tr (v_{xx}\Sigma\Sigma^{\top})+v_x^{\top}f+q -\frac{1}{2}v_x^{\top}\Sigma\Gamma R^{-1}\Gamma^{\top}\Sigma^{\top}v_x=0, \quad (t,x)\in [0,T)\times \R^n,\\&v(T,x)=g(x), \quad x\in \R^n,
		\end{aligned}
		\right .
	\end{equation}
	which satisfies the format of (\ref{Cauchy}) with 
	\begin{align}\label{bf}
		b(t,x)&\equiv f(t,x)\\
		h(t,x,z)&\equiv q(t,x) -\frac{1}{2}z^{\top}\Gamma(t,x)R^{-1}\Gamma^{\top}(t,x)z.
	\end{align} 
	We may thus obtain the (viscosity) solution of (\ref{HJB3}) by simulating the system of FBSDE given by (\ref{fsde}) and (\ref{bsde}). Notice that (\ref{fsde}) corresponds to the uncontrolled ($u=0$) system dynamics.  
	
\section{Approximating the Solution of FBSDEs}\label{Sec:Num}
The solution of FBSDEs has been studied to a great extent independently from its connection to PDEs, mainly within the field of mathematical finance. Though several generic schemes exist \cite{Bouchard2004,Bender2007,Lemor2006}, in this paper we propose a scheme which exploits the regularity present in FBSDEs that arise from the application of the nonlinear Feynman-Kac lemma.

We begin by selecting a time grid $ \{t=t_0<\ldots<t_N=T\}$ for the interval $[t,T]$, and denote by $\Dti\triangleq t_{i+1}-t_i$ the $(i+1)$-th interval of the grid (which can be selected to be constant) and $\DWi\triangleq W_{t_{i+1}}-W_{t_i}$ the $(i+1)$-th Brownian motion increment\footnote{Here, $\DWi$ would be simulated as $\sqrt{\Dti}\xi_i$, where $\xi_i \sim \mathcal{N}(0,I)$.}. For notational brevity, we also denote $X_i\triangleq X_{t_i}$. The simplest discretized scheme for the forward process is the Euler scheme, which is also called \textit{Euler-Maruyama} scheme \cite{Kloeden1999}:
\begin{equation}\label{EulMar}
\left \{
\begin{aligned}
&X_{i+1}\approx X_i+b(t_i,X_i)\Dti + \Sigma(t_i,X_i)\DWi,\\
&i=1,\ldots,N, \qquad X_0=x.
\end{aligned}
\right.
\end{equation}
Several alternative, higher order schemes exist that can be selected in lieu of the Euler scheme\cite{Kloeden1999}.
To discretize the backward process, we further introduce the notation $Y_i\triangleq\Yti$ and $Z_i\triangleq\Zti$. Then, recalling that adapted BSDE solutions impose $Y_s\triangleq\E[Y_s|\F_s]$ and $Z_s\triangleq\E[Z_s|\F_s]$ (i.e., a back-propagation of the conditional expectations), we approximate equation (\ref{bsde}) by
\begin{equation}\label{scheme1}
\small{Y_i=\E[Y_i|\F _{t_i}]\approx \E[Y_{i+1}+h(t_{i+1},X_{i+1},Y_{i+1},Z_{i+1})\Dti|X_i].}
\end{equation}
Notice that in the last equality the term $Z^{\top}_i\DWi$ in (\ref{bsde}) vanishes because of the conditional expectation ($\DWi$ is zero mean), and we replace $\F _{t_i}$ with $X_i$ in light of the Markovian property presented in Section \ref{FBSDEpresent}.
By virtue of equation (\ref{FKbsde}), the $Z$-process in (\ref{bsde}) corresponds to the term $\Sigma^{\top}(s,\Xstx)v_x(s,\Xstx)$. Therefore we can write
\begin{align}
Z_i&=\E[Z_i|\F_{t_i}]=\E[\Sigma^{\top}(t_i,X_i )\nabla_xv(t_i,X_i)|X_i] \nonumber\\
&= \Sigma^{\top}(t_i,X_i)\nabla_xv(t_i,X_i),
\end{align} 
which naturally requires knowledge of the solution at time $t_i$ on a neighborhood $x$, $v(t_i,x)$.
The backpropagation is initialized at $Y_T=g(X_T)$ and $Z_T=\Sigma(T, X_T)^{\top}\nabla_xg(X_T)$,
for a $g(\cdot)$ which is differentiable almost everywhere. There are several ways to approximate the conditional expectation in (\ref{scheme1}), however in this work we shall employ the Least Squares Monte Carlo (LSMC) method\footnote{Treating conditional expectations by means of linear regression was made popular in the field of mathematical finance by \cite{Longstaff2001}.}, which we shall briefly review in what follows. 

The LSMC method addresses the general problem of numerically estimating conditional expectations of the form $\E[Y|X]$ for square integrable random variables $X$ and $Y$, if one is able to sample $M$ independent copies of pairs $(X,Y)$. The method itself is based on the principle that the conditional expectation of a random variable can be modeled as a function of the variable on which it is conditioned on, that is, $\E[Y|X]=\phi^*(X)$, where $\phi^*$ solves the infinite dimensional minimization problem
\begin{equation}\label{LSMCproj}
\phi^*=\arg\min_{\phi}\E[|\phi(X)-Y|^2],
\end{equation}
and $\phi$ ranges over all measurable functions with $\E[|\phi(X)|^2]<\infty$. A finite-dimensional approximation of this problem can be obtained if one decomposes $\phi(\cdot)=\sum_{i=1}^{k}\varphi_i(\cdot)\alpha_i=\varphi(\cdot)\alpha$, with $\varphi(\cdot)$ being a row vector of predetermined basis functions and $\alpha$ a column vector of constants,  thus solving $\alpha^*=\arg\min_{\alpha \in \R^k}\E[|\varphi(X)\alpha-Y|^2]$, with $k$ being the dimension of the basis. Finally, this problem can be simplified to a linear least-squares problem if one substitutes the expectation operator with its empirical estimator \cite{Gyoerfi2002}, thus obtaining
\begin{equation}\label{LSMCEE}
\alpha^*=\arg\min_{\alpha \in \R^k}\frac{1}{M}\sum_{j=1}^{M}|\varphi(X^j)\alpha-Y^j|^2,
\end{equation}
wherein $(X^j,Y^j)$, $j=1,\ldots,M$ are independent copies of $(X,Y)$. Introducing the notation
\begin{equation}\label{basisf}
\Phi(X)=\begin{bmatrix}
\varphi(X^1)\\
\vdots\\
\varphi(X^M)
\end{bmatrix}\in \R^{M\times k},
\end{equation}
the solution to this least-squares problem can be obtained by directly solving the normal equation, i.e.,
\begin{equation}
\begin{aligned}
\alpha^*&=\bigg( \Phi^\top(X)\Phi(X)\bigg)^{-1}\Phi^\top(X)\bigg(\begin{bmatrix}
Y^1\\
\vdots\\
Y^M
\end{bmatrix}\bigg),
\end{aligned}
\end{equation} 
or by performing gradient descent. The LSMC estimator for the conditional expectation assumes then the form $\E[Y|X=x]=\phi^*(x)\approx \varphi(x)\alpha^*$.

Returning to our problem, we may apply the LSMC method to approximate the conditional expectation in equation (\ref{scheme1}) for each time step. To this end, we require a  vector of basis functions $\varphi$ for the approximation of $\E[Y_i|X_i]$. Although the basis functions can be different at each time step, we shall use the same symbol for notational simplicity. Then, Monte Carlo simulation is performed  by sampling $M$ independent trajectories $\{X^m_i\}_{i=1,\ldots,N}$, in which the index $m=1,\ldots,M$ specifies a particular Monte Carlo trajectory. Whenever this index is not present, the entirety with respect to this index is to be understood. The numerical scheme is initialized at the terminal time $T$ and is iterated backwards along the entire time grid, until the starting time instant has been reached. At each time step $t_i$, we are given $M$ pairs of data $(Y^m_{i},X^m_{i})$\footnote{Here, $Y_i^m$ denotes the quantity $Y^m_{i+1} +\Dti h(t_{i+1},X^m_{i+1},Y^m_{i+1},Z^m_{i+1})$, which is the $Y^m_i$ sample value before the conditional expectation operator has been applied.} on which we perform linear regression to estimate the conditional expectation of $Y_i$ as a function of $x$ at the time step $t_i$. This provides us an approximation of the Value function $v$ at time $t_i$ for the neighborhood of the state space that has been explored by the sample trajectories at that time instant, since $v(t_i,x)=\E[Y_i|X_i=x]\approx\varphi(x)\alpha_{i}$. We then replace $Y^m_i=\E[Y^m_i|X^m_i]\approx\varphi(X_i^m)\alpha_{i}$, thereby treating the conditional expectation as a projection operator. Finally, the approximation of the conditional expectation of $Z_i$ is obtained by taking the gradient with respect to $x$ on $v(t_i,x)$, evaluating it at $X^m_i$, and scaling it with $\Sigma$
\begin{equation}
Z^m_i\approx\Sigma(t_i, X^m_i)^{\top}\nabla_x\varphi(X^m_i)\alpha_{i}.
\end{equation}
This process is repeated for $t_{i-1},\ldots,t_1$. Note that this approach requires the basis functions $\varphi(\cdot)$ of our choice to be differentiable almost everywhere, so that $\nabla_x\varphi(x)$ is available in analytical form for almost any $x$. The proposed algorithm is then summarized as
\begin{equation*}\label{ay}
 \left \{
 \begin{aligned}
&\mathrm{Initialize:} ~Y_T=g(X_T), \quad Z_T=\Sigma(T, X_T)^{\top}\nabla_xg(X_T),\\
&\alpha_{i}=\arg \min_{\alpha} \frac{1}{M}\Big \|
\Phi(X_i) \alpha - 
\bigg(Y_{i+1} +\Dti h(t_{i+1},X_{i+1},Y_{i+1},Z_{i+1})\bigg)
\Big \|^2,  \\
& Y_i=\Phi(X_i)\alpha_{i},\quad Z^m_i=\Sigma(t_i, X^m_i)^{\top}\nabla_x\varphi(X^m_i)\alpha_{i},  
 \end{aligned}
 \right .
\end{equation*} 
where $m=1,\ldots,M$ and the matrix $\Phi$ defined in (\ref{basisf}). Again, the minimizer 
 can be obtained by directly solving the normal equation, i.e.,
\begin{equation}
\begin{aligned}
\alpha_i&=\bigg( \Phi^\top(X_i)\Phi(X_i)\bigg)^{-1}\Phi^\top(X_i)\bigg( Y_{i+1}+\Dti h(t_{i+1},X_{i+1},Y_{i+1},Z_{i+1})\bigg),
\end{aligned}
\end{equation} 
or by performing gradient descent. The essential algorithm output is the collection of $\alpha_i$'s, that is, the basis function coefficients at each time instant, which are needed to recover the Value function approximation for the particular area of the state space that is explored by the forward process. This is in contrast with methods that calculate the solution over an entire pre-specified grid (and thus typically exhibit bad scalability), but also differs from local trajectory optimization methods which consider only infinitesimal variations around a nominal trajectory. Furthermore, an important difference between the proposed method and forward sampling based methods is that the latter provide a solution only for the point of the initial condition $(t,x)$, while the solution of this method covers an area starting from the initial condition $(t,x)$, expanding in state space until time $T$ is reached.  

\section{Learning Control: An Iterative Scheme based on Importance Sampling}
	
The proposed method, as it has been presented so far, suffers from a significant limitation. Namely, its ability to provide approximations to the value function is restricted to only those areas of the state space that are reachable by unforced dynamics (eq. (\ref{fsde})). Indeed, there are several cases of systems in which the goal state practically cannot be reached by the uncontrolled system dynamics (consider, for example, an inverted pendulum). Furthermore, even in the case in which the target state is indeed reached by unforced trajectories, as the dimensionality of the state space increases, the density of sample trajectories along any given path from the initial state to the target state reduces quickly, thus increasing the demand for available samples. These issues can be eliminated if one is given the ability to modify the drift term of the sampled trajectories. Specifically, by changing the drift, we can direct the exploration of the state space towards the goal state, or any other state of interest, reachable by control.
As will be shortly demonstrated, such a scheme can indeed be constructed by means of a careful application of Girsanov's theorem on the change of measure. Applying importance sampling on FBSDEs is not an entirely new concept, as it was first introduced as a variance reduction technique \cite{Moseler2010}. Through Girsanov's theorem, one may alter the drift of the forward process if this modification is appropriately compensated for in the backward process. That is, the system of FBSDEs given by equations (\ref{fsde}) and (\ref{bsde}) is in some certain sense equivalent to one with modified drift
\begin{equation}\label{stdyn3}
\left \{
\begin{aligned}
&\dd \Xt_s=[b(s,\Xt_s)+\Sigma(s,\Xt_s)K_s]\dd s + \Sigma(s,\Xt_s)\dd W_s, \qquad s\in[t,T],\\&\Xt_t=x.
\end{aligned}
\right.    
\end{equation}
along with the compensated BSDE
\begin{equation}\label{Ydyn3}
\left \{
\begin{aligned}
&\dd \Yt_s=[-h(s,\Xt_s,\Yt_s,\Zt_s)+\Zt_s^{\top}K_s]\dd s +\Zt_s^{\top}\dd W_s,\qquad s\in[t,T],\\ &\Yt_T=g(\Xt_T),
\end{aligned}
\right.
\end{equation}
for any measurable, bounded and adapted process $K:[0,T]\to \R^p$. Equivalence is not path-wise of course, since the paths realized by both the forward and the backward processes will be different under the modified drift dynamics. However, the solution at starting time $t$, that is $(Y_t,Z_t)$, will remain unaffected. In other words, the estimate of the Value function at the initial condition $(t,x)$ is independent of the drift term modification, as will be proven shortly. Indeed, following Girsanov's Theorem \cite{Karatzas1991,Oksendal2007}, we define a new measure $\probb$ with $\dd \probb (\omega)=M(T;t,\omega)\dd \prob(\omega)$, where
\begin{equation*}
M_s\triangleq\exp\bigg(-\int_t^sK_\tau^{\top}\dd W_\tau-\frac{1}{2}\int_t^s|K_\tau|^2\dd \tau \bigg), \quad s\in[t,T],
\end{equation*}
is the process of Radon-Nikodym derivatives $\dd \probb^{(s)}/\dd \prob^{(s)}$ with $\probb^{(s)}$ and $\prob^{(s)}$ being the restrictions of $\probb$ and $\prob$ to $\F_s$, respectively. Then, $M_s$ is a $\prob$-martingale, the $\prob$-law of $(X,Y,Z)$ is the same as the $\probb$-law of $(\Xt,\Yt,\Zt)$, and $$\tilde{W}_s\triangleq\int_{t}^sK_\tau\dd \tau+W_s, \qquad s\in [t,T],$$
is a Brownian motion under $\probb$. In fact, defining the $\probb$-Brownian increment $\dd \tilde{W}_s=K_s\dd t+\dd W_s$, it becomes evident that equations (\ref{stdyn3}) and (\ref{Ydyn3}) are simply copies of the dynamics of equations (\ref{fsde}) and (\ref{bsde}), if one substitutes $\dd W_s$  in the latter with  $\dd \tilde{W}_s$. Now, notice that since at the time of initialization, $t$, $M_t$ is by construction equal to one with probability one (in both $\prob$ and $\probb$-measure), the measures $\prob$ and $\probb$ restricted to $\F_t$ are equal, and therefore the pairs  $(Y_t,Z_t)$ and  $(\Yt_t,\Zt_t)$ are equal in expectation as well. This proves that the Value function at the initial condition $(t,x)$ is independent of the drift term modification. An additional intuitive, albeit informal, explanation of why the modified system of FBSDEs (\ref{stdyn3}), (\ref{Ydyn3}) can be used in lieu of the original FBSDE system is readily obtained if one examines the associated PDEs. Indeed, the FBSDE problem defined by (\ref{stdyn3}) and (\ref{Ydyn3}) corresponds to the PDE problem
\begin{equation*}\label{HJBImpSamp}
	\left \{
	\begin{aligned}
		&v_t+\frac{1}{2}\tr (v_{xx}\Sigma\Sigma^{\top})+ v_x^{\top}(b+\Sigma K) +h(t,x,v,\Sigma^{\top}v_x)-v_x^{\top}\Sigma K =0, \quad (t,x)\in [0,T)\times \R^n,\\&v(T,x)=g(x),
	\end{aligned}
	\right .
\end{equation*}
which of course is identical to the PDE problem (\ref{Cauchy}), as we have merely added and subtracted the term $v_x^{\top}\Sigma K$. Thus, although the FBSDEs are different, they are associated with the same PDE problem. 

Returning to the original problem formulation and recalling the definition of $\Gamma(\cdot)$ in Assumption \ref{GSigma}, we may apply any nominal control $\bar{u}$ to the state dynamics in order to obtain the modified drift system, which exhibits the form
\begin{equation}
\small{\dd x(t)=[f(t,x(t))+\Sigma(t,x(t))\Gamma(t,x(t))\bar{u}(t)]\dd t+\Sigma(t,x(t))\dd W_t.}
\end{equation}
Thus, the controlled system trajectories are samples from the forward process (\ref{stdyn3}) with
\begin{equation}
K_s=\Gamma(s,X_s)\bar{u}(s), \qquad s\in[t,T],
\end{equation}
while $b(s,X_s)\equiv f(s,X_s)$ as per (\ref{bf}).  Notice that the nominal control $\bar{u}$ may be any open or closed-loop control, a random control, or even a control calculated by a previous run of the algorithm. In the latter case, one obtains a more refined solution, thus arriving at an iterative scheme. For the discrete representation on the time grid of Section \ref{Sec:Num}, we define $K_i=K_{t_i}$. The forward process can again be sampled using the Euler-Maruyama scheme. There are several equivalent ways to incorporate importance sampling in the backward process, however the most straightforward way is to simply define 
\begin{equation}\label{htilde}
\tilde{h}(s,x,y,z,k)\triangleq h(s,x,y,z)-z^{\top}k,
\end{equation}
and utilize the discretized scheme presented in Section \ref{Sec:Num} using $\tilde{h}$ instead of $h$. 
	\section{Simulation Results} \label{Sim}
	To evaluate the algorithm's performance, 
	we simulated the algorithm on an inverted pendulum and a cart-pole system. These simulations demonstrate that the nonlinearity in the dynamics is handled efficiently, and furthermore illustrate the significance of the iterative scheme which features importance sampling.

\subsection{The Inverted Pendulum}
\begin{figure}[htb]
		\centering
			\includegraphics[width=0.6\textwidth,height=0.35\textwidth]{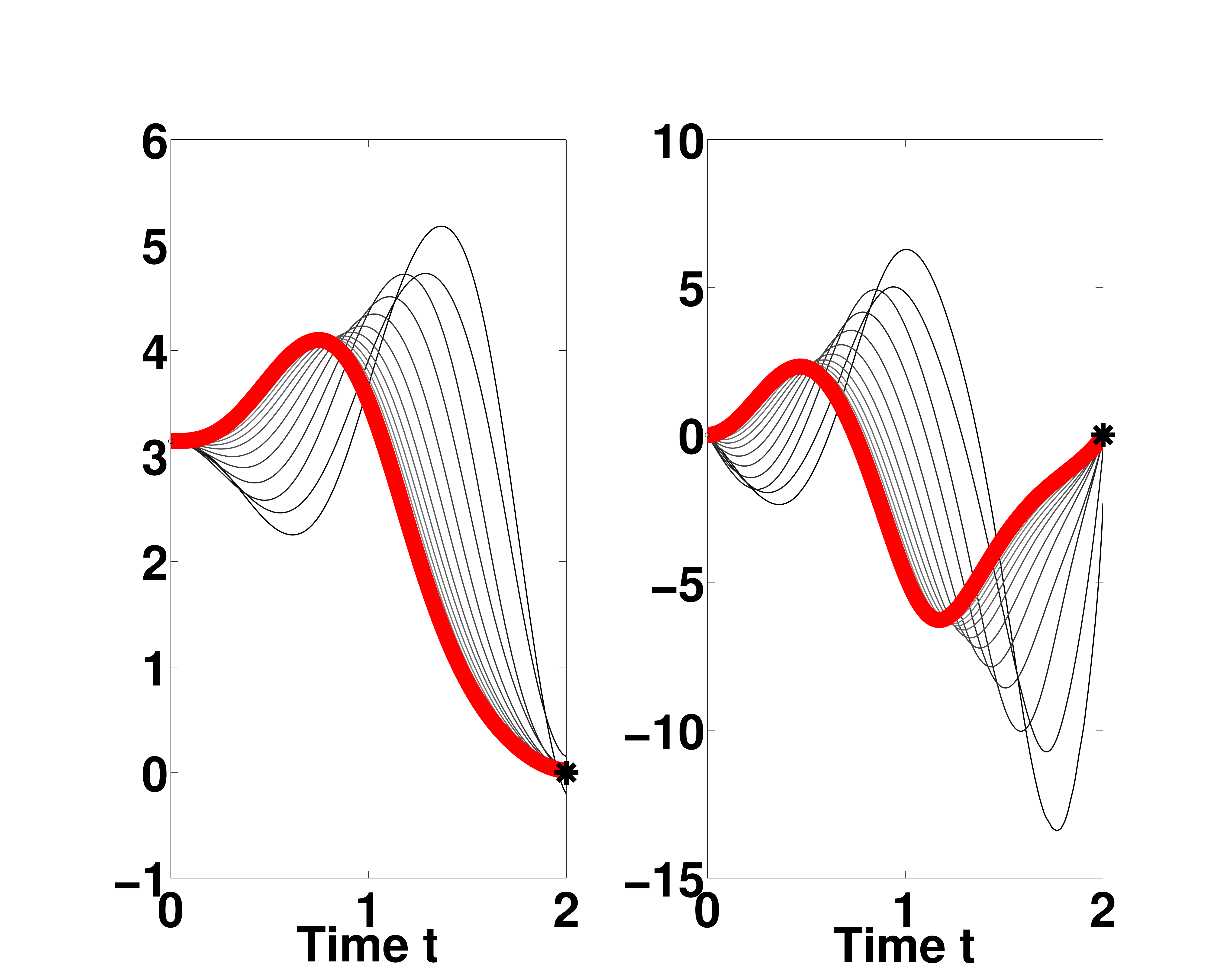}
		\caption{\small Inverted Pendulum: Trajectory mean for the position (left) and velocity (right) of the controlled system for each iteration (gray scale) and after the final iteration (red). The black dots represent the target states. }		\label{fig:pend}
	\end{figure}

\begin{figure}[htb]
		\centering
			\includegraphics[width=0.45\linewidth]{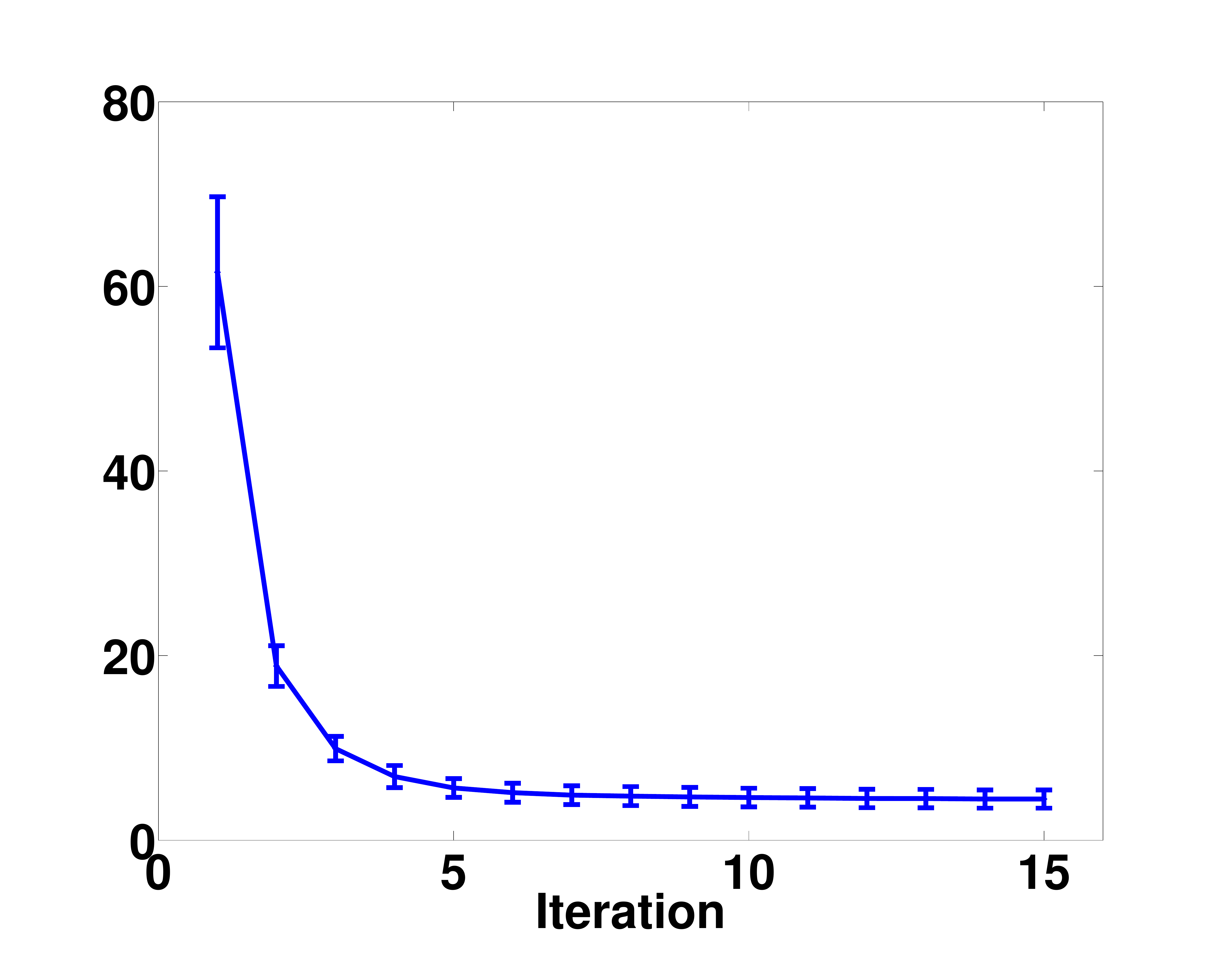}
			\vspace{-0.5cm}
				\caption{\small Inverted Pendulum:  Cost mean +- 3 standard deviations per iteration.}
				\label{fig:pend_cost}
	     \end{figure}

	The equations of motion for the inverted pendulum are given by
	\begin{equation}
	m\ell^2\ddot{\theta}+b\dot{\theta}-mg\ell\sin\theta=u,
	\end{equation}
		and stochasticity enters the system in form of perturbations in the torque $u$. For the purposes of this simulation, two thousand trajectories were generated on a time grid of 0.004 with time horizon T=2. The system noise covariance was set on 0.1. No initial guess for the control input was necessary, though a white noise signal has been injected in the control input during the sampling stages to increase variation in the trajectories, since the system noise intensity is low. For the basis of the Value function approximation, modified Chebyshev polynomials \cite{King1984} up to second order have been selected. The scheme was repeated for 15 iterations, with the algorithm successfully learning the optimal control to invert and stabilize the pendulum. Figure \ref{fig:pend} depicts 
		the mean of the controlled trajectories for each algorithm iteration (gray scale). The trajectories after the final iteration are shown in red. 
		Finally, Figure \ref{fig:pend_cost} depicts the convergence of the cost mean and standard deviation as the iterative scheme progresses.
	
	     \vspace{-0.6cm}

	\subsection{The Cart-Pole system}
\begin{figure}[htb]
		\centering
			\includegraphics[width=0.4\linewidth]{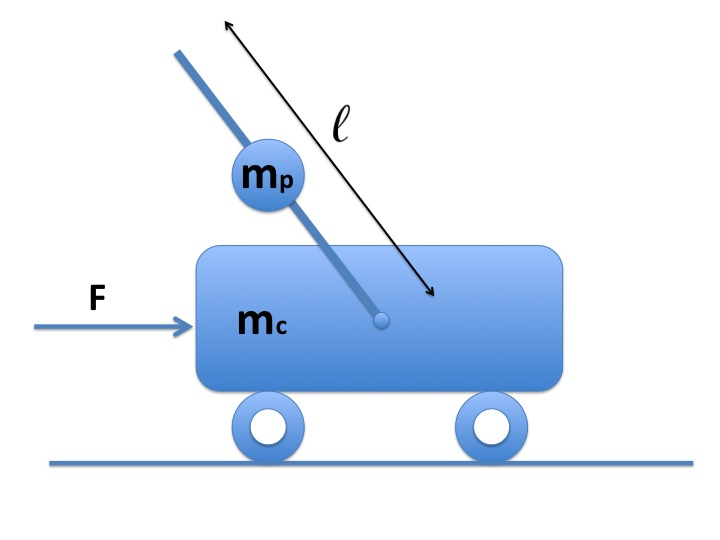}
				\caption{\small  Cart pole:  $m_{c} $ denoted the mass of the cart, $ m_{p} $ denotes the mass of the pole and $ \ell $ is the length of the pole.}
				\label{fig:Cart_Pole}
	     \end{figure}

\begin{figure}[htb]
		\centering	
			\includegraphics[width=0.5\linewidth]{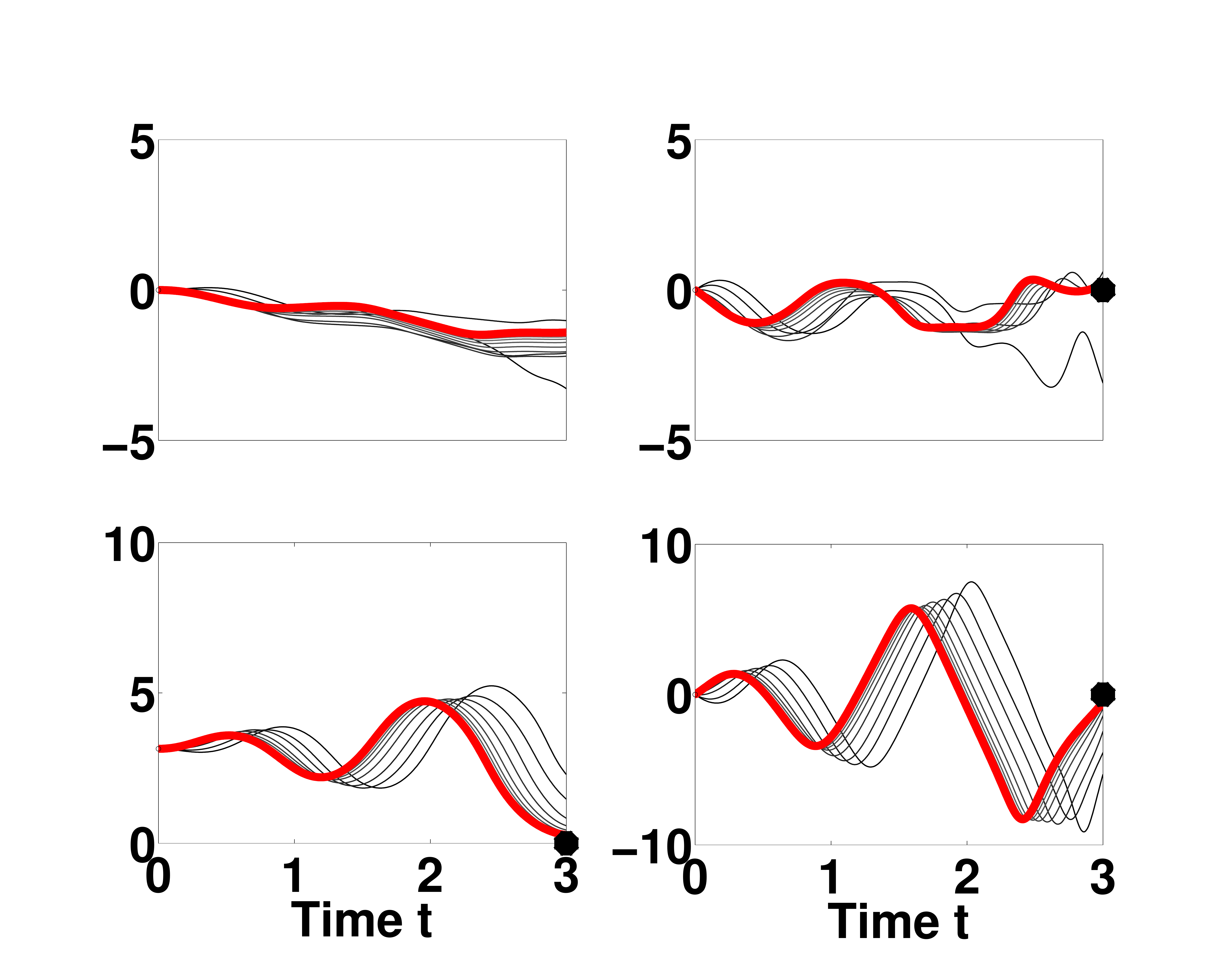}
			\vspace{-0.6cm}
			\caption{\small Cart-pole: Clockwise starting at the top left-- cart position, cart velocity, pole velocity, pole position. Trajectory mean of the controlled system for each iteration (gray scale) and after the final iteration (red). The black dots represent the target states. }
		\label{fig:cp}
	\end{figure}

\begin{figure}[htb]
		\centering	
			\includegraphics[width=0.4\linewidth]{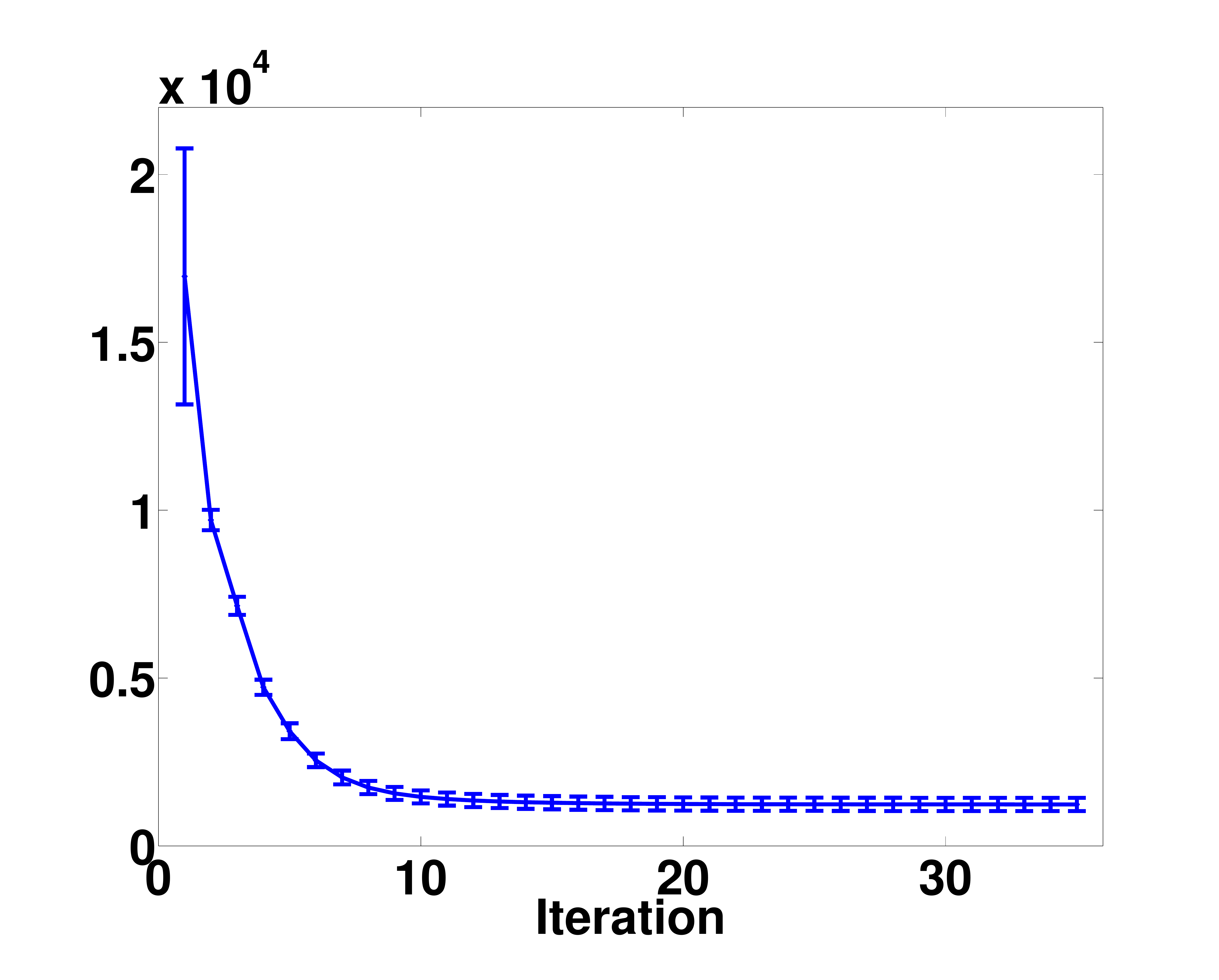}
			\vspace{-0.5cm}
		\caption{\small Cart-pole:  Cost mean +- 3 standard deviations per iteration.}
			\label{fig:cp_cost}
	\end{figure}

	To assess the efficiency of the proposed scheme in underactuated systems, we simulated the algorithm on a cart-pole system (see Figure \ref{fig:Cart_Pole}). The equations of motion are given by
	\begin{align}
		\ddot{x}&=\frac{1}{m_c+m_p\sin^2\theta}\bigg(u-m_p\sin\theta(\ell\dot{\theta}^2+g\cos\theta)\bigg),\\
		\ddot{\theta}&=	\frac{1}{\ell(m_c+m_p\sin^2\theta)}	\bigg(u\cos\theta-m_p\ell\dot{\theta}^2\cos\theta\sin\theta +(m_c+m_p) g\sin\theta\bigg),
	\end{align}
	and stochasticity enters the system in form of perturbations in $u$. To this end, five thousand trajectories were generated on a time grid of 0.004 with time horizon T=3. The system noise covariance was set on 1. Again, no initial guess for the control input was necessary, and a white noise signal has been injected in the control input during the sampling stages to increase variation in the trajectories, since the system noise intensity is low. For the basis of the Value function approximation, modified Chebyshev polynomials up to second order have been selected. The scheme was repeated for 35 iterations. Figure \ref{fig:cp} depicts 
	the mean of the controlled trajectories for each algorithm iteration (gray scale). The trajectories after the final iteration are shown in red. 
	Finally, Figure \ref{fig:cp_cost} depicts the convergence of the cost mean and standard deviation as the iterative scheme progresses.
	     
	
	\section{Conclusions} \label{Conc}
	In this paper we proposed a new algorithm for nonlinear stochastic control problems with dynamics affine in control and cost functions that are non-quadratic in the state and quadratic in controls. In light of a nonlinear Feynman-Kac lemma which establishes a connection between certain PDEs and SDEs, we were able to obtain a probabilistic representation of the solution to the nonlinear HJB PDE, expressed as a system of FBSDEs. This system is then simulated using linear regression. Finally, in order to enhance the algorithm's efficiency in treating more complex nonlinear systems, we proposed an iterative scheme based on Girsanov's theorem on the change of measure, which features importance sampling. We demonstrated the ability of the proposed iterative algorithm to learn the optimal controls without an initial guess in an inverted pendulum system and a cart-pole system.
	
	\section*{Acknowledgment}
	
The first author gratefully acknowledges support from the A. S. Onassis Foundation.


\bibliographystyle{ieeetr}
\bibliography{stochastics}

\begin{thebibliography}{10}

\bibitem{Mitchell2003}
I.~M. Mitchell and C.~J. Tomlin, ``Overapproximating reachable sets by
  {Hamilton-Jacobi} projections,'' {\em Journal of Scientific Computing},
  vol.~19, no.~1-3, pp.~323--346, 2003.

\bibitem{Aguilar2014}
C.~O. Aguilar and A.~J. Krener, ``Numerical solutions to the {Bellman} equation
  of optimal control,'' {\em Journal of Optimization Theory and Applications},
  vol.~160, no.~2, pp.~527--552, 2014.

\bibitem{Horowitz2014}
M.~B. Horowitz and J.~W. Burdick, ``Semidefinite relaxations for stochastic
  optimal control policies,'' in {\em American Controls Conference (ACC)},
  pp.~3006--3012, 2014.

\bibitem{Horowitz2014a}
M.~B. Horowitz, A.~Damle, and J.~W. Burdick, ``Linear {Hamilton} {Jacobi}
  {Belman} equations in high dimensions,'' in {\em 53rd IEEE Conference on
  Decision and Control, Los Angeles, California, USA}, December 15-17 2014.

\bibitem{Gorodetsky2015}
A.~Gorodetsky, S.~Karaman, and Y.~Marzouk, ``Efficient high-dimensional
  stochastic optimal motion control using tensor-train decomposition,'' in {\em
  Robotics: Science and Systems (RSS)}, 2015.

\bibitem{Todorov2005}
E.~Todorov and W.~Li, ``A generalized iterative {LQG} method for locally
  optimal feedback control of constrained nonlinear stochastic systems,'' {\em
  American Control Conference}, pp.~300--306, 2005.

\bibitem{Theodorou2010}
E.~A. Theodorou, Y.~Tassa, and E.~Todorov, ``Stochastic differential dynamic
  programming,'' {\em American Control Conference}, pp.~1125--1132, 2010.

\bibitem{Kappen2005}
H.~J. Kappen, ``Linear theory for control of nonlinear stochastic systems,''
  {\em Physical Review Letters}, vol.~95, November 2005.

\bibitem{Theodorou2010a}
E.~A. Theodorou, J.~Buchli, and S.~Schaal, ``A generalized path integral
  control approach to reinforcement learning,'' {\em The Journal of Machine
  Learning Research}, vol.~11, pp.~3137--3181, January 2010.

\bibitem{Dvijotham2012}
K.~Dvijotham and E.~Todorov, ``Linearly solvable optimal control,'' {\em
  Reinforcement Learning and Approximate Dynamic Programming for Feedback
  Control}, pp.~119--141, 2012.

\bibitem{Todorov2009}
E.~Todorov, ``Efficient computation of optimal actions,'' {\em Proceedings of
  the {National Academy of Sciences}}, vol.~106, no.~28, pp.~11478--11483,
  2009.

\bibitem{Theodorou2015}
E.~A. Theodorou, ``Nonlinear stochastic control and information theoretic
  dualities: Connections, interdependencies and thermodynamic
  interpretations,'' {\em Entropy}, vol.~17, no.~5, pp.~3352--3375, 2015.

\bibitem{Karatzas1991}
I.~Karatzas and S.~Shreve, {\em Brownian Motion and Stochastic Calculus}.
\newblock Springer-Verlag New York Inc., 2nd~ed., 1991.

\bibitem{Kappen2005a}
H.~J. Kappen, ``Path integrals and symmetry breaking for optimal control
  theory,'' {\em Journal of Statistical Mechanics: Theory and Experiment},
  vol.~11, November 2005.

\bibitem{Yong1999}
J.~Yong and X.~Y. Zhou, {\em Stochastic Controls: Hamiltonian Systems and HJB
  Equations}.
\newblock Springer-Verlag New York Inc., 1999.

\bibitem{Fleming2006}
W.~Fleming and H.~Soner, {\em Controlled Markov Processes and Viscosity
  Solutions}.
\newblock Stochastic Modelling and Applied Probability, Springer, 2nd~ed.,
  2006.

\bibitem{Ma1999}
J.~Ma and J.~Yong, {\em Forward-Backward Stochastic Differential Equations and
  Their Applications}.
\newblock Springer-Verlag Berlin Heidelberg, 1999.

\bibitem{ElKaroui1997}
N.~El~Karoui, S.~Peng, and M.~C. Quenez, ``Backward stochastic differential
  equations in finance,'' {\em Mathematical Finance}, vol.~7, January 1997.

\bibitem{Bouchard2004}
B.~Bouchard and N.~Touzi, ``Discrete time approximation and {Monte Carlo}
  simulation of {BSDE}s,'' {\em Stochastic Processes and their Applications},
  vol.~111, pp.~175--206, June 2004.

\bibitem{Bender2007}
C.~Bender and R.~Denk, ``A forward scheme for backward {SDE}s,'' {\em
  Stochastic Processes and their Applications}, vol.~117, pp.~1793--1812,
  December 2007.

\bibitem{Lemor2006}
J.~P. Lemor, E.~Gobet, and X.~Warin, ``Rate of convergence of an empirical
  regression method for solving generalized backward stochastic differential
  equations,'' {\em Bernoulli}, vol.~12, no.~5, pp.~889--916, 2006.

\bibitem{Kloeden1999}
P.~Kloeden and E.~Platen, {\em Numerical Solution of Stochastic Differential
  Equations}, vol.~23 of {\em Applications in Mathematics, Stochastic Modelling
  and Applied Probability}.
\newblock Springer-Verlag Berlin Heidelberg, 3rd~ed., 1999.

\bibitem{Longstaff2001}
F.~A. Longstaff and R.~S. Schwartz, ``Valuing {American} options by simulation:
  A simple least-squares approach,'' {\em Review of Financial Studies},
  vol.~14, pp.~113--147, 2001.

\bibitem{Gyoerfi2002}
L.~Gy\"{o}rfi, M.~Kohler, A.~Krzyzak, and H.~Walk, {\em A Distribution-Free
  Theory of Nonparametric Regression}.
\newblock Springer Series in Statistics, Springer-Verlag New York, Inc., 2002.

\bibitem{Moseler2010}
T.~Moseler and C.~Bender, ``Importance sampling for backward {SDE}s,'' {\em
  Stochastic Analysis and Applications}, vol.~28, no.~2, pp.~226--253, 2010.

\bibitem{Oksendal2007}
B.~{\O}ksendal, {\em Stochastic Differential Equations- An Introduction with
  Applications}.
\newblock Springer-Verlag Berlin Heidelberg, 6th~ed., 2007.

\bibitem{King1984}
J.~T. King, {\em Introduction to Numerical Computation}.
\newblock McGraw-Hill, Inc., 1984.

\end{thebibliography}

\end{document}